\documentclass[article,10pt]{IEEEtran}
\onecolumn

\usepackage[table]{xcolor}
\usepackage[T1]{fontenc}

\usepackage[boxed,vlined,ruled]{algorithm2e}
\usepackage{booktabs}
\usepackage{lineno}
\usepackage{lipsum}
\usepackage{soul}
\usepackage{cite}
\usepackage{balance}
\usepackage{graphicx}
\usepackage{setspace}
\usepackage{enumitem}

\usepackage{array}
\usepackage{amsthm}
\usepackage{amssymb}
\usepackage{amsfonts}
\usepackage[cmex10]{amsmath}
\usepackage[amsmath]{empheq}
\usepackage{cases}
\usepackage{mathrsfs}
\usepackage{bbm}

\newtheorem{theorem}{Theorem}

\newtheorem{lemma}{Lemma}

\newtheorem{proposition}{Proposition}

\renewcommand{\qed}{\hfill$\blacksquare$}

\newcommand{\rem}[1]{}

\def\sr{\textsc{Sr-Adc}}
\def\tsr{\emph{self-reset}}
\def\tf{\emph{folding}}
\def\fo{f_{{\sf{Out}}}}
\def\fin{f_{\sf{In}}}

\def\Z{\in \mathbb{Z}}

\def\l{\left(}
\def\r{\right)}
\def\DE{\triangleq}
\def\ind{\mathbbmtt{1}}

\def\T{T}
\def\B{\beta_g}
\def\kn{\kappa_{\l n \r}}
\def\S{\mathsf{S}}

\def\sinc{\mathrm{sinc}}

\renewcommand\bar{\overline}

\newcommand{\fig}[1]{Fig.~\ref{#1}}
\newcommand{\md}[2]{#1~\rm{mod}~#2}

\newcommand{\MO}[1]{\mathscr{M}_\lambda\l #1 \r}
\newcommand{\VO}[1]{\varepsilon_{#1}}
\newcommand{\VOB}[1]{\bar{\varepsilon}_ {#1}}

\newcommand{\ft}[1]{\left[\kern-0.15em\left[#1  \right]\kern-0.15em\right]}
\newcommand{\fe}[1]{\left[\kern-0.30em\left[#1  \right]\kern-0.30em\right]}
\newcommand{\flr}[1]{\left\lfloor #1 \right\rfloor}

\newcommand{\DL}[1]{\stackrel{(\mathrm{#1})}{=}}
\newcommand{\EQc}[1]{\stackrel{(\ref{#1})}{=}}

\newcommand{\BL}[1]{#1 \in \mathcal{B}_{\Omega}}
\newcommand{\BLP}[1]{#1 \in \mathcal{B}_{\pi}}

\definecolor{b1}{RGB}{0,128,255}
\definecolor{b2}{RGB}{0,0,204}
\definecolor{r1}{RGB}{200,0,0}
\definecolor{p1}{RGB}{167,28,138}
\definecolor{p2}{RGB}{255,0,128}

\newcommand{\com}[1]{{\color{black} #1}}
\newcommand{\felix}[1]{{\color{black} #1}}

\newcommand{\abv}[1]{{\color{black}{#1}}}

\ifCLASSINFOpdf
\else
\fi
\begin{document}
\title{On Unlimited Sampling}


\author{
\vspace{30pt}
\IEEEauthorblockN{\bf Ayush Bhandari} \\
\IEEEauthorblockA{Massachusetts Institute of Technology\\
77 Mass. Ave. Cambridge, 02139 USA\\
Email: \texttt{ayush@MIT.edu}}\\

\vspace{30pt}

\and

\IEEEauthorblockN{\bf Felix Krahmer}\\
\IEEEauthorblockA{Technical University of Munich\\
Boltzmannstra{\ss}e 3, 85747 Garching/Munich, Germany\\
Email: \texttt{felix.krahmer@tum.de}}\\

\vspace{30pt}

\and

\IEEEauthorblockN{\bf Ramesh Raskar}\\
\IEEEauthorblockA{Massachusetts Institute of Technology\\
77 Mass. Ave. Cambridge, 02139 USA\\
Email: \texttt{raskar@MIT.edu}}

\thanks{Initial version of the full paper that will appear in the Proceedings of the 12$^{\text{th}}$ International Conference on Sampling Theory and Applications, July 3--7, 2017, Tallinn, Estonia.}
}


%


\maketitle

\vfill
{\color{black}
\begin{abstract}
Shannon's sampling theorem provides a link between the continuous and the discrete realms stating that bandlimited signals are uniquely determined by its values on a discrete set. 
This theorem is realized in practice using so called analog--to--digital converters (ADCs). Unlike Shannon's sampling theorem, the ADCs are limited in dynamic range. Whenever a signal exceeds some preset threshold, the ADC saturates, resulting in aliasing due to clipping. The goal of this paper is to analyze an alternative approach that does not suffer from these problems. Our work is based on recent developments in ADC design, which allow for ADCs that reset rather than to saturate, thus producing modulo samples. An open problem that remains is: Given such modulo samples of a bandlimited function as well as the dynamic range of the ADC, how can the original signal be recovered and what are the sufficient conditions that guarantee perfect recovery? In this paper, we prove such sufficiency conditions and complement them with a stable recovery algorithm. Our results not limited to certain amplitude ranges, in fact even the same circuit architecture allows for the recovery of arbitrary large amplitudes as long as some estimate of the signal norm is available when recovering. Numerical experiments that corroborate our theory indeed show that it is possible to perfectly recover function that takes values that are orders of magnitude higher than the ADC's threshold.
\end{abstract}}
\IEEEpeerreviewmaketitle

\newpage
\tableofcontents

\newpage

\linespread{2.5}
\section{Introduction}

There is no doubt that Shannon's sampling theorem is one of the cornerstone results in signal processing, communications and approximation theory. An informal statement of this theorem is that any function with compactly supported Fourier domain representation can be uniquely represented from a set of equidistant points taken on the function. The spacing between the points---the sampling rate---is a function of the bandwidth. In the last few decades, the field has grown far and wide \cite{Butzer:1992,Zayed:1993,Unser:2000,Butzer:2014} and a number of elegant extensions have been reported. These extensions relate to (but not limited to) shift-invariant subspaces \cite{Boor:1994,Blu:1999,Aldroubi:2001,Eldar:2009}, union of subspaces \cite{Lu:2008,Mishali:2011}, sparse signals \cite{Vetterli:2002,Blu:2008,Adcock:2015}, phase space representations \cite{Bhandari:2012}, spectrum blind sampling \cite{Feng:1996} and operator sampling \cite{Pfander:2013,Krahmer:2014}. Broadly speaking and in most cases, the variation on the theme of sampling theory arises from the diversity along the \emph{time-dimension}. For example, sparsity vs.~smoothness and uniform vs.~non-uniform grid are attributes based on the time dimension. A variation on the hypothesis linked with any of these attributes leads to a new sampling theorem. Here, we take a different approach and discuss a hypotheses based on the \emph{amplitude-dimension}. 

\begin{figure}[!b]
\label{fig:adc}
\centering
\includegraphics[width =0.6\textwidth]{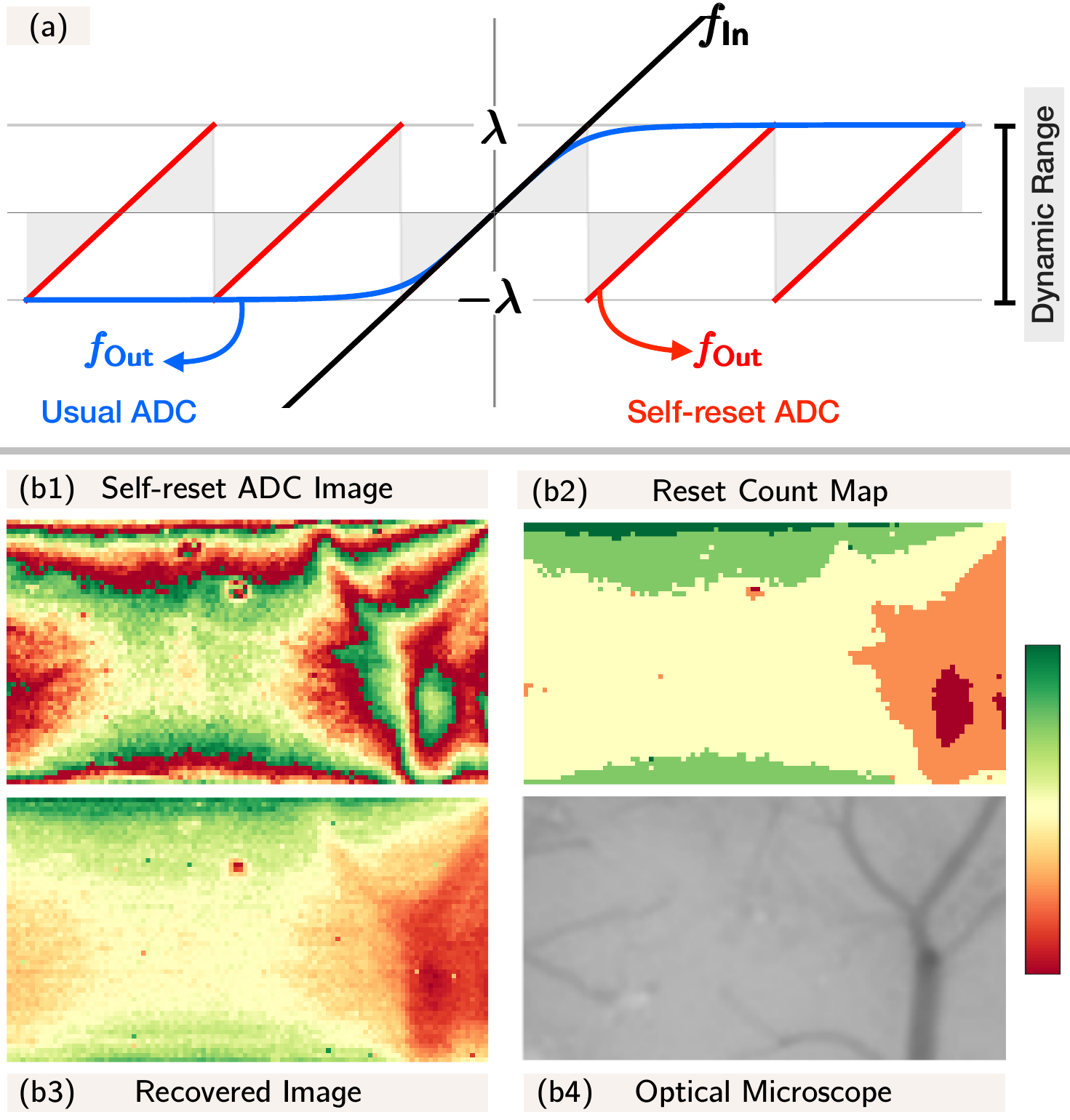}
\caption{Usual ADC compared with self-reset ADC. (a) Whenever the input signal $\fin$ voltage exceeds a certain threshold $\lambda$, the output signal $\fo$ in any conventional ADC saturates to $\lambda$ and this results in clipping. In contrast, whenever $| \fin |>\lambda$, the self-reset ADC folds $\fin$ such that $\fo$ is always in the range $\left[-\lambda,\lambda\right]$. In this way, the self-reset configuration circumvents clipping but introduces discontinuities. (b) Images obtained with prototype self-reset ADC. (b1) Image obtained with a self-reset ADC shows folded amplitudes. (b2) For each pixel, the ``reset count map'' shows the number of times the image amplitude has undergone folding. (b3) Unfolded image based on reset count map. (b4) Image obtained using an optical microscope.}
 \end{figure}

\subsection{Shannon's Sampling Theory and a Practical Bottleneck}
From a practical standpoint, point-wise samples of the function are obtained using so-called analog-to-digital converter (or the ADC). \rem{This where the topic of \emph{quantization} finds its way to the sampling theory. Quantization, in context of sampling theory of band-limited signals, is a mature topic which is fairly well understood.} That said, one bottleneck that inhibits the functionality of an ADC is its \emph{dynamic range}---the maximum recordable voltage range $\left[-\lambda,\lambda\right]$. Whenever a signal crosses this threshold, the ADC saturates and the resulting signal is clipped. This is shown in \fig{fig:adc}-(a). 

The issue of clipping is a serious problem and manifests as non-linear artifacts in audio-visual data as well as applications that involve sampling of physiological or bio-medical data. Not surprisingly, a number of works have studied this problem in different contexts (cf.~\cite{Abel:1991,Olofsson:2005,Adler:2011,Ting:2013}). When a bandlimited signal is clipped, the resulting function has discontinuities which amounts to creating high-frequency distortions to the signal. Hence, if a signal is not amplitude limited, it may be prone to aliasing artifacts \cite{Esqueda:2016}. Although important, this aspect is rarely discussed in context of Shannon's sampling theory \cite{Zayed:1993,Unser:2000,Butzer:2014,Boor:1994,Blu:1999,Aldroubi:2001,Eldar:2009}---something that is very relevant to our work. Clipped signals are typically handled by restoration algorithms \cite{Logan:1984,Esqueda:2016,Ting:2013} which seek to recover permanently lost data samples under certain assumptions.

\subsection{A Solution via Modular Arithmetic}

\begin{figure*}[!t]
\centering
\includegraphics[width =1\textwidth]{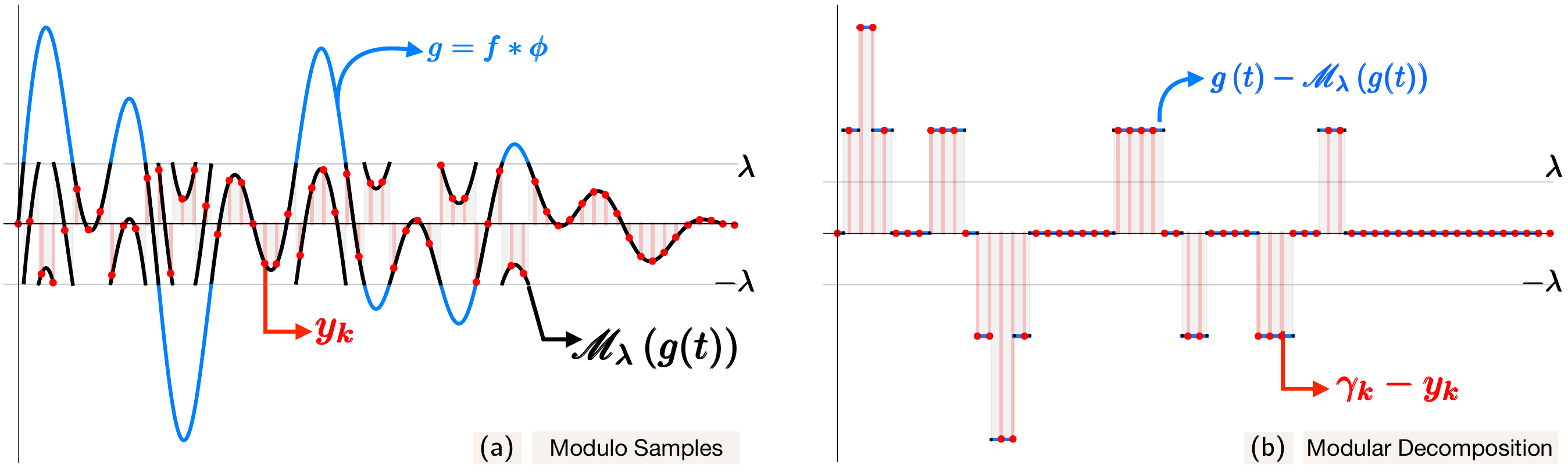}
\caption{(a) For a prototype low-pass filtered data $g = f*\phi$ (in {\color{b1} {\bf ------}}) we plot the continuous version of the modulo-ADC~$\MO{g\l t \r}$ (in {\color{black} {\bf ------}}) together with uniform samples $y_n$ (in {\color{r1} {\bf ---$\bullet$}}). (b) We plot the continuous time residual $\VO{g}\l t \r = g\l t \r-\MO{g\l t \r}$ (in {\color{b1} {\bf ------}}) and its sampled version $\{\VO{\gamma} = \gamma_k - y_k\}$ (in {\color{r1} {\bf ---$\bullet$}}).}
\label{fig:mod}
\end{figure*}
 
\begin{figure}[!h]
\centering
\includegraphics[width =0.5\textwidth]{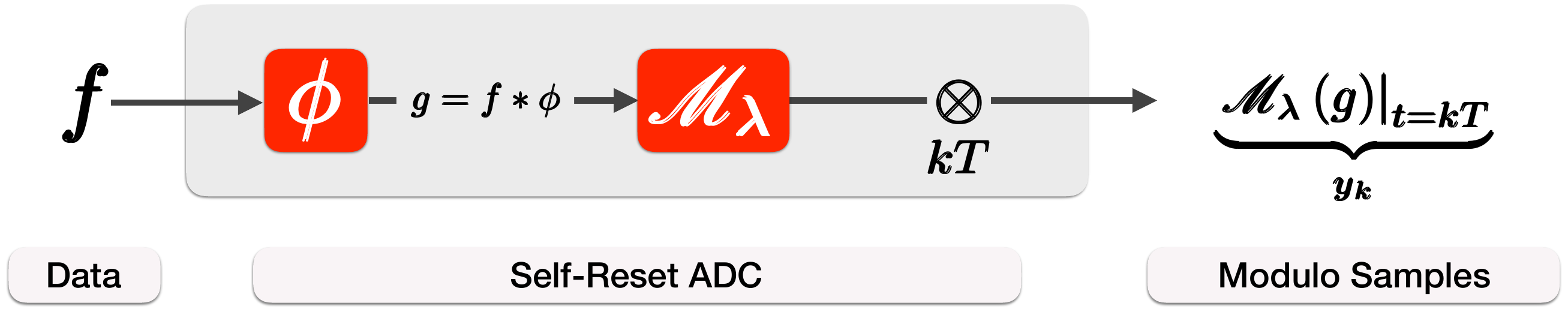}
\caption{\sr~architecture for obtaining modulo samples.}
\label{fig:blk}
\end{figure}



Thanks to recent advancements in ADC design, a radically different approach is rapidly developing. Namely, when reaching the upper or lower saturation threshold, these ADCs would reset to the respective other threshold, in this way allowing to capture subsequent changes even beyond the saturation limit.
Motivated by Tomlinson decoders as they appear in the context of communication theory \cite{Gopal:2015}, the theoretical concept of such ADCs already appeared in the literature as early as in the late 1970's under the name of a modulo limiter \cite{Ericson:1979, Chou:1992}, but only the nature of the resulting quantization noise, and neither their physical realization, nor their recovery properties were investigated. 


Physical realizations only started to develop in the early 2000's.
Depending on the community, the resulting ADC constructions are known as \tf{-}ADC (cf.~\cite{Kester:2009} and references therein) or the \tsr-ADC, recently proposed by Rhee and Joo \cite{Rhee:2003} in context of CMOS imagers \felix{(see \fig{fig:adc}-(b1-b3) for visualization of their approach)}. As noted in \cite{Kavusi:2004}, the \sr{s} allow for simultaneous enhancement of the dynamic range as well as the signal-to-noise ratio. The goal of developing \sr{s} is motivated by the fact that the dynamic range of natural images typically surpasses what can be handled by the usual ADCs. This feature is not only critical to consumer photography \cite{Kavusi:2004} but also plays an important role in life sciences and bio-imaging. For instance, last year, Sasagawa \cite{Sasagawa:2016} and Yamaguchi \cite{Yamaguchi:2016} developed \sr{s} for functional brain imaging.


Common to all such architectures \cite{Rhee:2003,Kavusi:2004,Chou:1992,Ericson:1979} is a memoryless, non-linear mapping of the form,
\begin{equation}
\label{map}
\mathscr{M}_{\lambda}:t \mapsto 2\lambda \left( {\fe{ {\frac{t}{{2\lambda }} + \frac{1}{2} } } - \frac{1}{2} } \right),
\end{equation}
where $\ft{t} \DE t - \flr{t} $ defines the fractional part of $t$. The mapping in (\ref{map}) is folding amplitudes, that is, the range of $\mathscr{M}_{\lambda}$ is $\left[ -\lambda,\lambda\right]$. To visualize this operation, we plot $\mathscr{M}_{\lambda}(t)$ as a function of $t$ in \fig{fig:adc}-(b). In effect, (\ref{map}) can be interpreted as a centered modulo operation since \felix{$\mathscr{M}_{\lambda}(t)\equiv \md{t}{2\lambda}$.}

%
%


In comparison to the remarkable progress that has been made on the hardware front of the new ADCs, theoretical and algorithmic aspects related to the identifiability and recovery properties of the modulo samples remain largely unexplored. \felix{Namely, all previous approaches \cite{Rhee:2003,Sasagawa:2016,Yamaguchi:2016} would require additional information such as a detailed account of how often the amplitudes have been folded \felix{(cf.~\fig{fig:adc}-(b2))}, which leads to significantly more involved circuit architectures as well as additional power and storage requirements. No attempts have been made to take the viewpoint of an inverse problem, that is, to identify or recover the signal from modulo samples alone.} In this paper, we aim to bridge this gap. 


\vspace{4pt}

\noindent Our contributions are two fold: 

\begin{enumerate}[leftmargin=15pt,label=$\arabic*)$, itemsep = 3pt]

\item We take a first step towards formalization of a sampling theorem which describes sufficient conditions for sampling and reconstruction of bandlimited functions in context of the \sr \ and modulo samples.

\item Our sufficiency condition is complemented by a constructive algorithm which describes a recovery principle and that is stable with respect to noise. 
\end{enumerate}
In this way, our work allows for \emph{unlimited sampling}---the case when the amplitude range of a function far exceeds the dynamic range of the usual ADCs, that is, $\max | \fin| \gg \lambda$. This is achieved by trading sampling rate for dynamic range. 


\section{Self-reset ADC and Modulo Samples}
\label{sec:SR}

%
%
Motivated by the electronic architectures discussed in \cite{Rhee:2003,Kester:2009}, we will use the model in \fig{fig:blk} for representing the \sr~using which we obtain modulo samples. 
%
%
Here $f\in L_2$ is the function to be sampled and $\phi$ is the sampling kernel. Let $\widehat \phi \left( \omega  \right) = \int {\phi \left( t \right){e^{ - \jmath \omega t}}dt}$ denote the Fourier Transform of $\phi$. We say $\phi$ is $\Omega$--bandlimited or,
\[
\BL{\phi} \Leftrightarrow \quad \widehat \phi \left( \omega  \right) = {\ind_{\left[ { - {\Omega},{\Omega}} \right]}}\left( \omega  \right)\widehat \phi \left( \omega  \right) \mbox{ and } \phi \in L_2
\]
where $\ind_{\mathcal{D}}\l t \r$ is the indicator function on domain $\mathcal{D}$. In practice, $f$ may not be bandlimited, in which case, pre-filtering with $\BL{\phi}$ ensures that the signal to be sampled is bandlimited. In the remainder of this paper, we will assume that we are given a low-pass filtered version of $f$, which we will refer to as $g \DE f*\phi$. Furthermore, we will normalize the bandwidth to $\pi$ such that $\BLP{g}$. This function $g$ then undergoes a non-linear, amplitude folding defined in (\ref{map}) and results in, 
\begin{equation}
\label{yc}
z\l t \r = \MO{g\l t \r}.
\end{equation}
Finally, the output of (\ref{map}), that is, $\mathscr{M}_{\lambda}(g(t))$ is sampled using impulse-modulation, \abv{${ \otimes _{k \T }} \DE \sum\nolimits_{n \in \mathbb{Z}} {\delta \left( {t - k \T }\right)}$}, where $\T>0$ is the sampling rate. This results in uniform samples, 
\begin{align}
\label{yn}
{y_k} & \DE z\left( {k \T } \right) 
 = \MO{g\l k\T \r}, \ \ k\Z  
\end{align}
as shown in \fig{fig:blk}. To develop a sense about the functionality of the \sr, in \fig{fig:mod}-(a), we plot $g\l t \r$, $z\l t \r $ and samples $y_k$. It is clear that the self-reset ADC converts a smooth function into a discontinuous one. This is an important aspect that is attributed to the presence of simple functions arising from (\ref{map}).


\begin{proposition}[Modular Decomposition]\label{prop:mod} Let $\BLP{g}$ be a zero-mean function and $\MO{g\l t \r}$ be defined in (\ref{map}) with $\lambda$ fixed, non-zero constants. Then, the bandlimited function $g$ admits a decomposition
%
\begin{equation}
\label{eg}
\felix{
g\l t \r =  z\l t \r +{{\varepsilon _g}\left( t \right)},} 
\end{equation}
where ${\varepsilon _g}$ is a simple function, ${\varepsilon _g}\left( t \right) = 2\lambda\sum\nolimits_{\ell \in {\mathbb{Z}}} {{e_\ell}{{\ind}_{{\mathcal{D}_\ell}}}\left( t \right)}, e_\ell \in \mathbb{Z}$.
\label{prp:MD}
\end{proposition}


The proof of this proposition is provided in the appendix. In \fig{fig:mod}-(b), we plot the sampled version of $\varepsilon_g$ corresponding to $g$ in \fig{fig:mod}-(a). What (\ref{eg}) shows is that recovering $g$ from $z$ boils down to finding $\varepsilon_g$. Next, we explore sufficiency conditions for recovery of $g$ given modulo-samples $\{y_k\}_k$.

\section{A Sufficiency Condition and a Sampling Theorem}

In this section, we concretely answer the following questions:
 
\noindent \emph{Let $\BLP{g}$ and suppose that we are given modulo samples of $g$ defined by $y_k$ in (\ref{yn}),
\begin{enumerate}[leftmargin=30pt,label=\rm{Q}${\arabic*}$:, itemsep = 3pt]
  \item What are the conditions for recovery of $g$ from $y_k$?
  \item In the spirit of Shannon's sampling theorem, is there a constructive algorithm that maps samples $y_k \to g$?
\end{enumerate}}
In what follows, we will answer the two questions affirmatively.

\subsection{A Sufficiency Condition for Unlimited Sampling}

Given the sequence of modulo samples $y_k$, our basic strategy will be to apply a higher order finite difference operator $\Delta^N$, where the first order finite difference $\Delta$ is given by $(\Delta y)_k = y_{k+1}-y_k$. We will be exploiting that such operators commute with the modulo operation. So after applying the amplitude folding \eqref{map} to the resulting sequence, one obtains the same output as if one had started with $g_k$ instead of $y_k$. That in turn will allow for recovery if the higher order finite differences of the $g_k$'s are so small that the amplitude folding has no effect.

Consequently, our goal will be to investigate when higher order finite differences of samples of a bandlimited function are small enough. One way to ensure this is to sufficiently oversample. A first step towards this goal will be to relate the sample finite difference and the derivative of a bounded function. This well-known observation is summarized in the following lemma, a proof of which is included \felix{in the appendix} for the reader's convenience.

\begin{lemma}
\label{Lemma}
For any $g\in C^m(\mathbb{R}) \cap L_\infty(\mathbb{R}) $, its samples $\gamma_k\DE g(kT)$ satisfy 
\begin{equation}
\label{lem:1}
\|\Delta^N \gamma\|_\infty \leq T^N e^N \| g^{(N)}\|_\infty.
\end{equation}
\end{lemma}

\noindent 
To bound the right hand side of (\ref{TI}), we invoke Bernstein's inequality (cf.~pg.~116 in \cite{Nikolskii:1975}),
\begin{equation}
\label{BI}
||{g^{\left( N \right)}}|{|_\infty } \leqslant {\pi ^N}||g|{|_\infty}.
\end{equation} 
Consequently, by combining (\ref{TI}) and (\ref{BI}), we obtain, 
\begin{equation}
\label{TIBI}
||{\Delta ^N}\gamma|{|_\infty } \leqslant {\left( {\T \pi e} \right)^N}||g|{|_\infty }.
\end{equation}

This inequality will be at the core of our proposed recovery methods. Namely, provided $T<\tfrac{1}{\pi e}$, choosing $N$ logarithmically in $\|g\|_\infty$ 
ensures that the right hand side of \eqref{TIBI} is less than $\lambda$. 
More precisely, assuming that some $\B\in 2\lambda \mathbb{Z}$ is known with $||g||_\infty\leqslant \B$, a suitable choice is 
\begin{equation}
\label{NB_set}
N = \left\lceil {\frac{{\log \lambda  - \log {\B }}}{{\log \left( {{T}\pi e} \right)}}} \right\rceil.
\end{equation}
For the remainder of this paper we will work with this choice of $N$ and assume $T\leq \tfrac{1}{2\pi e}$, which yields the assumption in a stable way. We believe that a more precise analysis along the same lines will also yield corresponding bounds for $T \in [\tfrac{1}{2\pi e},\tfrac{1}{\pi e})$.

The bound of $\lambda$ for \eqref{TIBI} in turn entails that the folding operation has no effect on $\Delta^N \gamma$, that is, 
\begin{equation}
\label{eq:modguar}
\bar{\gamma}\DE \Delta^N \gamma \equiv \mathscr{M}_{\lambda}(\bar{\gamma})= \mathscr{M}_{\lambda}(\bar{y}), 
\end{equation}
which allows for recovery of $\bar{\gamma}$, as the right hand side can be computed from the folded samples $y$. \felix{Here, the last equality in \eqref{eq:modguar} follows by applying the following standard observation, a proof of which is included in 
the appendix for the reader's convenience, to $a=\gamma-\tfrac{1}{2}$.}

\begin{proposition} 
\label{prp:COM}
For any sequence $a$ it holds that
\begin{equation}
\label{yngn}
\abv{\mathscr{M}_{\lambda}(\Delta^N a) =\mathscr{M}_{\lambda}(\Delta^N({\mathscr{M}_{\lambda}(a)}).} 
\end{equation}
\end{proposition}
To choose $N$, we need \com{some upper bound} $\B$ such that $\|g\|_\infty \leq \B$, which we assume to be available for the remainder of this paper. As it simplifies the presentation, we assume w.l.o.g. that $\B\in 2\lambda \mathbb{Z}$. Note that $\B$ is only needed for recovery; there are no limitations on $\B$ arising from the circuit architecture.

To recover the sequence $\gamma$, recall from Proposition~\ref{prop:mod} that $\varepsilon_\gamma \DE \gamma - y$, that is, the sampled version of $\varepsilon_g$ takes as values only multiples of $2\lambda$. As $y$ is observed, finding $\varepsilon_\gamma$ is equivalent to finding $\gamma$.
Noting that $\bar{\varepsilon}_\gamma \DE \Delta^N \l \gamma -y\r$ can be computed from the data (via \eqref{eq:modguar}), it remains to investigate how $\varepsilon_\gamma$ can be recovered from $\bar{\varepsilon}_\gamma = \Delta^N\varepsilon_\gamma$. Due to the massive restriction on the range of all $\Delta^n \varepsilon_\gamma$, this problem is considerably less ill-posed than the problem of recovering $\gamma$ from $\Delta^N \gamma$. In particular, repeatedly applying the summation operator $\S: (a_i)_{i=1}^\infty \mapsto (\sum_{i'=1}^i a_{i'})_{i=1}^\infty$, the inverse of the finite difference, is a stable procedure because \abv{in the implementation we can round to the nearest multiple of $2\lambda$ in every step}.

There still is, however, an ambiguity that needs to be resolved. Namely, the constant sequence is in the kernel of the first order finite difference, so its inverse can only be calculated up to a constant. Thus when computing $\Delta^{n-1} \varepsilon_\gamma$ from $\Delta^n \varepsilon_\gamma, n=1,\ldots, N$, the result can only be determined up to an integer multiple of the constant sequence $\ell$ with value $2\lambda$, that is, $$\Delta^{n-1} \varepsilon_\gamma = \S\Delta^{n} \varepsilon_\gamma +\kn\ell,$$ for some $\kn\Z$. For $n=1$, this ambiguity cannot be resolved, as adding multiples of $2\lambda$ to a function results in the same modulo samples. 
To resolve this ambiguity for $n>1$, we apply the summation operator a second time. Repeating the same argument, we obtain that
\begin{equation}
\Delta^{n-2} \varepsilon_\gamma = \S^2\Delta^{n} \varepsilon_\gamma +L\kn+\felix{\kappa_{(n-1)}\ell},\label{eq:unb}
\end{equation}
where $L= \S\ell =(2\lambda i)_{i=1}^\infty$. This now implies that 
\begin{align}
{\left( {{\S^2}{\Delta ^n}{\varepsilon _\gamma }} \right)}_{\felix{1}} & - {\left( {{\S^2}{\Delta ^n}{\varepsilon _\gamma }} \right)}_{\felix{J + 1}} \hfill \notag \\
& = {\left( {{\Delta ^{n - 2}}{\varepsilon _\gamma }} \right)}_{\felix{1}} - {\left( {{\Delta ^{n - 2}}{\varepsilon _\gamma }} \right)}_{\felix{J + 1}} \felix{+} 2\lambda\kn J \notag \hfill \\
& = {\left( {{\Delta ^{n - 2}}{\gamma }} \right)}_{\felix{1}} - {\left( {{\Delta ^{n - 2}}{\gamma }} \right)}_{\felix{J + 1}} \felix{-}{\left( {{\Delta ^{n - 2}}{y }} \right)}_{\felix{1}} - {\left( {{\Delta ^{n - 2}}{y }} \right)}_{\felix{J + 1}} \felix{+} 2\lambda\kn J \notag \hfill \\
&   \in    2\lambda\kn J + \left( {2{{\left( {T\pi e} \right)}^{n - 2}}\B + 2^{n-1}\lambda} \right){{\left[ { - 1,1} \right]}} \label{eq:cont} \hfill \\
&   \subseteq   2\lambda J{\left[ \kn - \tfrac{\B}{{\lambda J}}- \tfrac{2^{n-2}}{{ J}},\kn + \tfrac{\B}{{\lambda J}} + \tfrac{2^{n-2}}{{ J}} \right]} \notag
\end{align}
where \eqref{eq:cont} uses \eqref{TIBI} together with the fact that $\|y\|_\infty \leq \lambda$. As $n\leq N$, \eqref{NB_set} yields that
\begin{equation}
2^{n-1} \leq 2^{N-1} \leq \tfrac{\beta_g}{\lambda} ^{-\tfrac{1}{\log_2(T\pi e)}} \leq \tfrac{\beta_g}{\lambda}, 
\end{equation}
where the last step uses that $T\leq\tfrac{1}{2\pi e}$.
So one obtains
\[
{\left( {{\S^2}{\Delta ^n}{\varepsilon _\gamma }} \right)}_{\felix{1}} - {\left( {{\S^2}{\Delta ^n}{\varepsilon _\gamma }} \right)}_{\felix{J + 1}} \in 2\lambda J{\left[ \kn - \tfrac{3\B}{{2\lambda J}},\kn + \tfrac{3\B}{{2\lambda J}} \right]}
\]
and choosing $J =\tfrac{6\B}{\lambda}$ directly yields that 

\begin{equation}
\label{kappan}
\kn = \left\lfloor \frac{(\S^2\Delta^{n} \varepsilon_\gamma)_{1} - (\S^2 \Delta^{n} \varepsilon_\gamma)_{J + 1}}{8\B} + \frac{1}{2} \right\rfloor.
\end{equation}
With this last ingredient, we are now ready to formulate our recovery method, Algorithm~\ref{alg:ModSamp}. For this algorithm, the discussion in the previous paragraph yields the following recovery result, the main result of this paper.

\begin{algorithm}[!t]
\label{alg:ModSamp}
\KwData{\ \ ${y_k} = \mathscr{M}_{\lambda}(g(kT))$, $N\in\mathbb{N}$, and $2\lambda \mathbb Z \ni \B \geqslant {\left\| g \right\|}_\infty.$}
\KwResult{$\widetilde g \approx g$.}
\begin{enumerate}[leftmargin=*,label=$\arabic*)$]
\item Compute $\bar{y} = \Delta^N y$.
\item Compute $\bar{\varepsilon}_\gamma =  \mathscr{M}_{\lambda}({\bar{y}}) -\bar{y}$. Set $s_{\l 1 \r} = \bar{\varepsilon}_\gamma$.
\item for $n = 1:N-1$

\hspace{10pt} Compute $\kappa_{\l n \r}$ in \eqref{kappan}.

\hspace{10pt} $s_{\l n + 1 \r} = \S s_{\l n \r} - 2\lambda \kappa_{\l n \r}  $.


end 

\item $\widetilde \gamma = \S s_{\l N \r}$.

\item Compute $\widetilde g$ from $\widetilde \gamma $ via low-pass filter. 
\end{enumerate}
\caption{Recovery from Modulo Folded Samples}
\end{algorithm}

 \begin{figure*}[!t]
\centering
\includegraphics[width =1\textwidth]{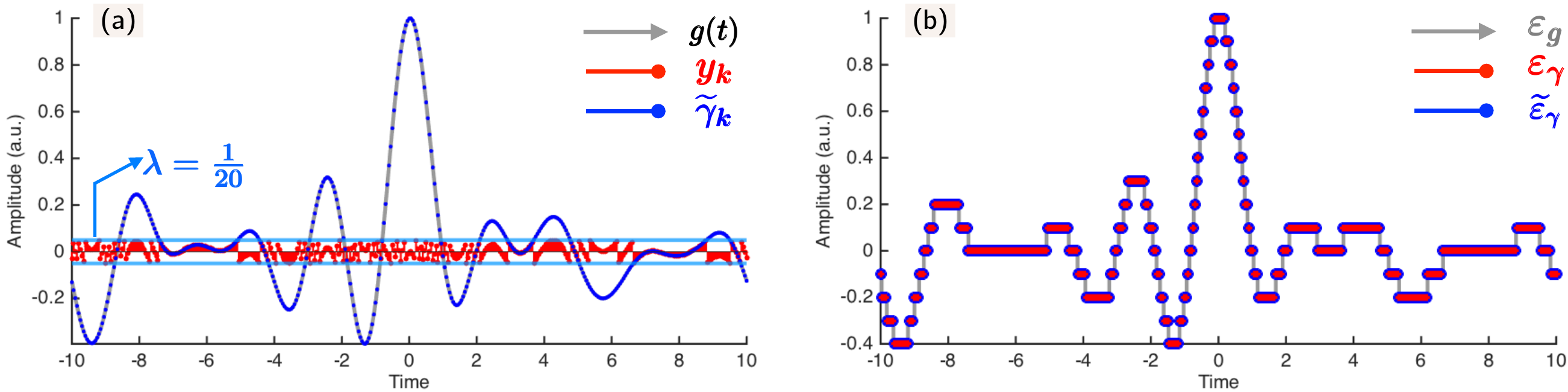}
\caption{Modulo sampling of bandlimited functions. (a) Randomly generated bandlimited function $\BLP{g}$, its modulo samples $y_k$ acquired with $\lambda = 1/20$ and $T \approx 1/2\pi e$, as well as reconstructed samples $\widetilde{\gamma}_k$ with $N = 3$. The mean squared error between ${\gamma_k} = g\left( {kT} \right)$ and reconstructed samples, $\widetilde{\gamma}_k$ was $1.6\times10^{-33}$. (b) We plot ${\varepsilon _g}\left( t \right) = g\left( t \right) - y\left( t \right)$ and compare the ground truth $\VO{\gamma}$ with its reconstructed version ${\widetilde \varepsilon _\gamma}$. The mean squared error between $\VO{\gamma}$ and $\widetilde{\varepsilon}_\gamma$ was of the order of machine precision, $1.5\times 10^{-33}$.}
\label{fig:result}
 \end{figure*}






%
%
\begin{theorem}[Unlimited Sampling Theorem] Let $\BLP{g\l t \r}$ and 
${y_k} = {\left. {\MO{ {g\left( t \right)}}} \right|_{t = kT}},k \in \mathbb{Z}$ in (\ref{yn}) be the modulo samples of $g\l t \r$ with sampling rate $\T$. Then a sufficient condition for recovery of $g\l t \r$ from the $\{y_k\}_k$ \abv{up to additive multiples of $2\lambda$} is that
\begin{equation}
\label{MSB}
\T \leq \frac{1}{2\pi e}.
\end{equation}
\noindent Provided that this condition is met and assuming that $\B\in 2\lambda \mathbb{Z}$ is known with $||g||_\infty\leqslant \B$, then choosing
\begin{equation}
\label{NB}
N = \left\lceil {\frac{{\log \lambda  - \log {\B }}}{{\log \left( {{T}\pi e} \right)}}} \right\rceil,
\end{equation}
yields that ${\left( {T\pi e} \right)^N}{\left\| {g} \right\|_\infty } < \lambda$ and Algorithm 1 recovers $g$ from $y$ \abv{again up to the ambiguity of adding multiples of $2\lambda$}. 
\end{theorem}

To put the theorem into perspective, note that in \eqref{MSB}, there are two degrees of freedom that are of a very different nature. On the one hand, the time $\T>0$ between two samples is intrinsically related to the circuit design, on the other hand, the number $N$ of finite difference applied is only relevant for the reconstruction. Hence only then, a bound for the supremum norm of the signal is required, the ADC itself is truly unlimited in that $T$ does not depend on $g$ and hence a fixed architecture allows to capture signals of arbitrary amplitude.

Recalling from the above discussion that one is in fact trying to satisfy the sufficient condition
\begin{equation}
\label{MSB}
{{\left( {T\pi e} \right)^N}{\left\| {g} \right\|_\infty } < \lambda},
\end{equation}
one sees that an alternative way to achieve this will be to consider a fixed $N$ and increase the oversampling rate. While this may be preferred in certain scenarios for computational reasons, this approach will no longer be unlimited, as decreasing $T$ would require to change the circuit architecture. That is why we decided to focus fixed $T$ and variable $N$ in our presentation.

\subsection{Numerical Demonstration}
We validate our approach via a numerical demonstration. We generate $\BLP{g}$ by multiplying the Fourier spectrum of the $\sinc$--function with weights drawn from the standard uniform distribution and then re-scale $g$ so that $||g||_\infty = 1$. With $\lambda = 1/20$ and $\T \approx 1/2\pi e$, we sample the signal and record $y_k$. This is shown in \fig{fig:result}(a). By implementing Algorithm~1 described above with $N=3$, rounding to the nearest multiple of $2\lambda$ for stability, we reconstruct samples $\widetilde{\gamma}_k$ as well as ${\widetilde \varepsilon _g}$ as shown in \fig{fig:result}(a). The reconstruction error is of the order of Matlab's numerical precision. For comparison purposes, we plot ${\varepsilon _g}$ and ${\widetilde \varepsilon _g}$ in \fig{fig:result}(b).

\section{Conclusions}

In this paper, we investigated sampling theory in context of a novel class of ADCs, so-called folding or self-reset ADCs. Using modulo operation, such ADCs fold amplitudes which exceed the dynamic range. Our main result establishes a sufficient condition when such folded samples allow for the recovery of bandlimited signals and a recovery method, thus providing a first step towards a sampling theory for self-reset ADCs. 

It should be noted that empirically, our sufficient condition typically does not seem to be optimal. Thus we consider it an interesting direction for further research to investigate sharper recovery conditions. Other natural extensions of our work relate to the problem of sampling in shift-invariant spaces. Furthermore, in many imaging problems, it is of interest to sample parametric signals. All such cases are important facets of modulo sampling theory which are completely unexplored.

\appendix
\section{Appendix (Proofs)}

\subsection{Proof of Proposition~\ref{prp:MD}}
\label{SecPrf:md}
Since $z\l t \r  = \MO{g\l t \r}$, by definition, we write, 
\begin{align}
 {\varepsilon _g}\left( t \right) & \EQc{eg}   g\l t \r - \MO{g\l t \r} \notag \\ 
& \DL{a} 2\lambda \left(h\l t \r- \ft{h\l t \r} \right) \DL{b} 2\lambda \flr{h\l t \r}, 
\label{egh}
\end{align}
where (a) is due to $\l g/2\lambda\r + 1/2 \mapsto h$ and in (b), we use $h = \ft{h} + \flr{h}$. Since $\flr{h}$, for an arbitrary function $h$, can be written as, 
\begin{equation}
\label{sf}
\flr{h\l t \r} = \sum\nolimits_{\ell \in \mathbb{Z}} {{e_\ell}{{\ind}_{{\mathcal{D}_\ell}}}\left(t \right)}, \quad e_\ell \in \mathbb{Z},
\end{equation}
{we obtain} the desired result. 
\qed

\subsection{Proof of Lemma~\ref{Lemma}}
\label{SecPrf:1}
Using Taylor's theorem, we can express $g\l t \r$ locally around an anchor point $\tau$ as the sum of a polynomial of degree $N-1$ and a remainder term of the form $r_l=\tfrac{g^{\l N \r}(\xi_l)}{N!}(l T-\tau)^N$ for some intermediate value $\xi_l$ between $lT$ and $\tau$. As $(\Delta^N \gamma)_k$ is a linear combination of $g\l t_n \r$'s with 
$$t_n=(k+n)T\in [kT, (n+k)T], \ \  n=0,\dots,N,$$
we choose the anchor point $\tau=(k+\tfrac{N}{2})T$. 
As $\Delta^N$ annihilates any polynomial of degree $N-1$, only the remainder term takes effect and we have
\[
(\Delta^N \gamma)_k = (\Delta^N r)_k = (\Delta^N r^{[k]})_0,
\]
where $r^{[k]}=(r_k, r_{k+1}, \dots, r_{k+N})$. Noting that for any vector $V$ one has by definition  $\|\Delta v\|_\infty \leq \|v\|_\infty$, it follows that
\begin{align}
\label{TI}
||{\Delta ^N}\gamma |{|_\infty } & \leqslant {2^N}||{r^{\left[ k \right]}}|{|_\infty } \notag \\
& \leqslant {2^N}\frac{{||{g^{\left( N \right)}}|{|_\infty }}}{{N!}}{\left( {\frac{{NT}}{2}} \right)^N}  \notag \\ 
& \leqslant {T^N}{e^N}||{g^{\left( N \right)}}|{|_\infty },
\end{align}
where in the last step, we used Stirling's approximation.
\qed

\subsection{Proof of Proposition~\ref{prp:COM}}
\label{SecPrf:com}
As usual, let $\bar{a} \DE \Delta^N a$. In view of Proposition~\ref{prp:MD} and (\ref{eg}), $a$ admits a unique decomposition, $a = \MO{a} + \VO{a}$ where $\VO{a}$ is a simple function. This allows us to write, $\Delta^N \MO{a}= \bar{a} - \VOB{a}$. Based on the distributive property of $\ft{t} \equiv \md{t}{1}$, it is not difficult to show that, $$\MO{a_1 + a_2} = \MO{\MO{a_1} + \MO{a_2}}.$$
And hence, 
\[
\MO{\Delta^N{\MO{a}}} = \MO{\bar{a} - \VOB{a}} = \MO{\MO{\bar{a}} - \MO{\VOB{a}} }.
\]
Now since $\VO{a}$ take values $2k\lambda$, $k\Z$, $\Delta^N \VO{a} = \VOB{a}$ takes values of the form $\{(-1)^n2\lambda\mathrm{B}^N_n\}_{n=0}^{N}\Z$ where $\mathrm{B}^N_n$ is the Binomial coefficient. From this, we conclude that $\VOB{a}$ is in the kernel of $\MO{\cdot}$ or $\MO{\VOB{a}} = 0$ and it follows that, 
$$\MO{\Delta^N{\MO{a}}} = \MO{\MO{\bar{a}}}= \MO{\bar{a}},$$
which proves the result in \eqref{yngn}.
\qed

\linespread{1}
\bibliographystyle{IEEEtran}
\bibliography{IEEEabrv,samptaRefs}
\end{document}